\documentclass[12pt]{article}

\usepackage[utf8]{inputenc}
\usepackage{hyperref}
\usepackage{geometry}
\usepackage{mathrsfs}
\usepackage{amsfonts}
\usepackage{amsmath}
\usepackage{amssymb}
\usepackage{bm}
\usepackage{enumerate}
\usepackage{mathtools}
\usepackage{units}
\usepackage{xcolor}
\usepackage{ulem}
\usepackage{graphics,graphicx}
\usepackage{float}

\title{Tunable phonon-driven magnon-magnon entanglement at room temperature}

\author{
    Yuefei Liu$^1$, Anders Bergman$^2$, Andrey Bagrov$^3$, Anna Delin$^{1,4}$, \\
    Danny Thonig$^{2,5}$, Manuel Pereiro$^2$, Olle Eriksson$^2$, \\
    Simon Streib$^2$, Erik Sj\"oqvist$^2$, Vahid Azimi-Mousolou$^{6,2}$
}

\date{November 2023}

\begin{document}

\maketitle

\noindent
$^1$ Department of Applied Physics, School of Engineering Sciences, KTH Royal Institute of Technology, AlbaNova University Center, SE-10691 Stockholm, Sweden \\
$^2$ Department of Physics and Astronomy, Uppsala University, Box 516, SE-751 20 Uppsala, Sweden \\
$^3$ Institute for Molecules and Materials, Radboud University, Heyendaalseweg 135, 6525AJ Nijmegen, The Netherlands \\
$^4$ Swedish e-Science Research Center (SeRC), KTH Royal Institute of Technology, SE-10044 Stockholm, Sweden \\
$^5$ School of Science and Technology, \"Orebro University, SE-701 82, \"Orebro, Sweden \\
$^6$ Department of Applied Mathematics and Computer Science, Faculty of Mathematics and Statistics, University of Isfahan, Isfahan 81746-73441, Iran

\newpage

\begin{abstract}
    We report the existence of entangled steady-states in bipartite quantum magnonic systems at elevated temperatures. We consider dissipative dynamics of two magnon modes in a bipartite antiferromagnet, subjected to interaction with a phonon mode and an external rotating magnetic field. To quantify the bipartite magnon-magnon entanglement, we use entanglement negativity and compute its dependence on temperature and magnetic field. We provide evidence that the coupling between magnon and phonon modes is necessary for the entanglement, and that, for any given phonon frequency and magnon-phonon coupling rate, there are always ranges of the magnetic field amplitudes and frequencies for which magnon-magnon entanglement persists at room temperature.
\end{abstract}


\section{introduction}

Entanglement is a central concept of quantum physics that has developed into a key resource in quantum technology \cite{browne2017}. It plays a role in quantum information processing, opens a way towards intercept-resilient quantum communications, and enables increased sensitivity in quantum metrology. If it were possible to overcome the thermal noise and prepare robust entangled quantum states at ambient conditions, the corresponding implications for the future quantum devices would be significant \cite{galve2010, dolde2013, wang2017, lin2020, zhong2020, arrazola2021, gulka2021, bello2022}. 

Magnons exhibit quantum properties over a wide range of temperatures and offer a possibility to access quantum phenomena at room temperature \cite{barman2021,yuan2022}. Compatibility of magnonics with hybrid quantum platforms \cite{yuan2022, Jie2018, toklikishvili2023, TABUCHI2016, Dany2019} makes magnons potentially useful for quantum data processing.  Most recently, particular attention has been paid to antiferromagnetic quantum magnonics due to its large equilibrium squeezing and entanglement \cite{yuan2022,bossini19,kamra2019,yuan2020,wuhrer22,Kamra2020,Yuan2021, Rezende2019,Azimi-mousolou2020,Azimi-mousolou2021,Azimi-mousolou2023,Shiranzaei2023}.

Here, we report evidence for equilibrium phonon-driven magnon-magnon entanglement in bipartite antiferromagnetic materials. We compute entanglement negativity \cite{Vidal2002} of a pair of magnon modes coupled to a lattice vibration, in the presence of an off-resonant external rotating magnetic field.
Our aim is to study the stability of bipartite entanglement between magnons to thermal noise. We search for the regime where the magnonic degrees of freedom remain entangled with each other rather than with the environment, and hence can be used as coherent channels in quantum information protocols.
We obtain a steady-state entanglement diagram in the temperature and magnetic field plane, which shows the possibility of having magnon-magnon entanglement at room temperature.
We analyze the magnon entanglement stability against dissipation caused by magnon-phonon coupling and observe that the two-mode magnon system indeed permits high-temperature entanglement at any value of the magnetic field provided that the magnetic field frequency is adjusted to the proper range. 
In turn, the existence of crystalline and synthetic antiferromagnetic materials, such as NiO \cite{Haakon2019}, 2D Ising systems like MnPSe$_3$ \cite{Thuc2021, Sheng2021},  YIG-based synthetic antiferromagets (SAFs) \cite{Changting2021}, and perovskite manganites \cite{Rini2007, Ulbrich2011}, particularly
SrTcO3 \cite{Rodriguez2011}, CaTcO3 \cite{Avdeev2011}, BiFeO3 \cite{Palkar2002} that have high Neel temperatures, provide a space for further study of high-temperature entanglement in quantum magnonics. 

\section{Physical system} 

\begin{figure}[h]
\centering
\includegraphics[width=.5\textwidth]{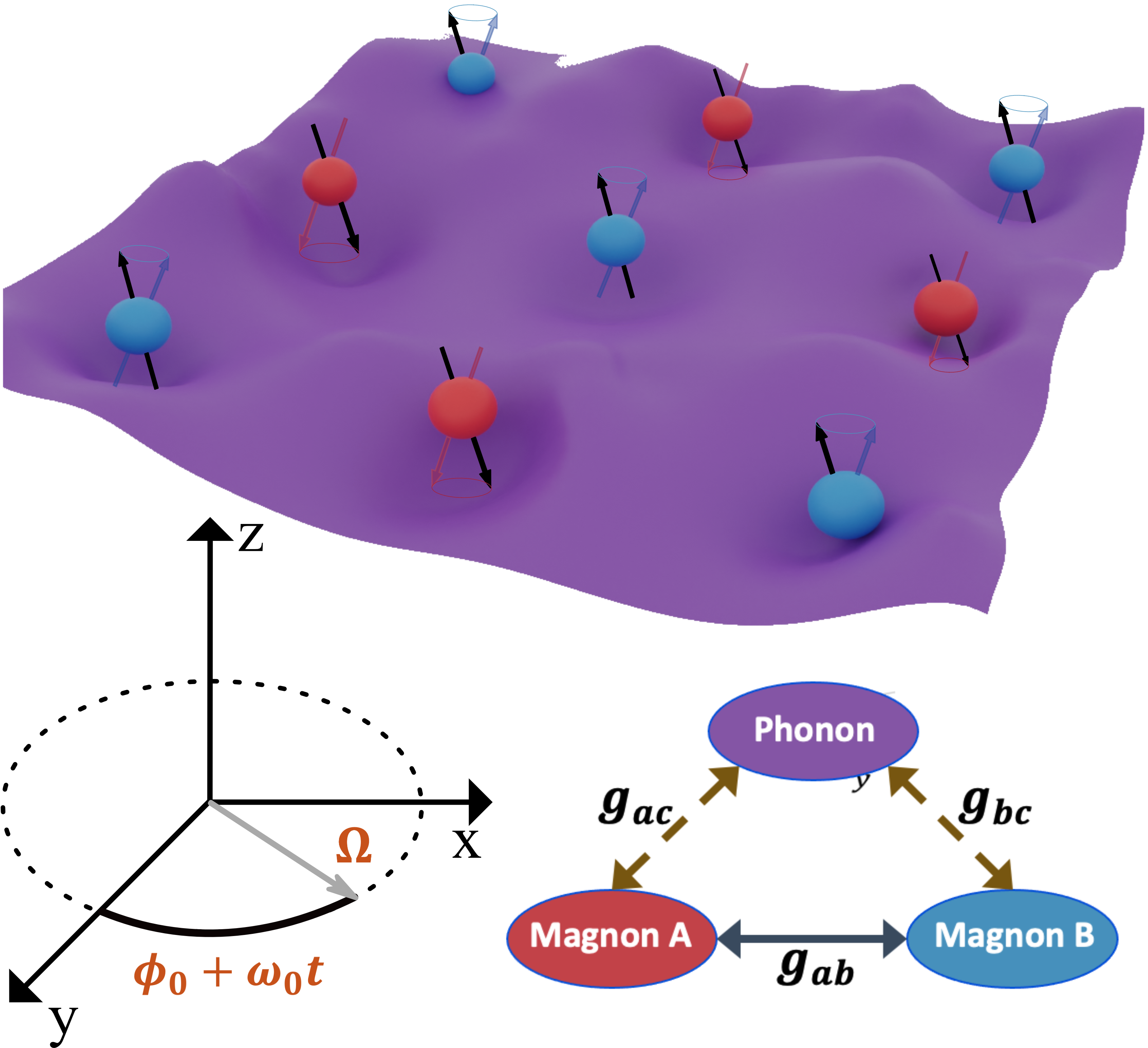} 
\caption{\label{fig:illus} (Color online) Schematic illustration of bipartite collective spin waves in an antiferromagnetic material subjected to a vibrational phonon mode in an off-resonant rotating magnetic field. The rotating magnetic field is illustrated in the lower left corner while the coupling between magnons in sublattice $A$ and $B$ to the phonons are illustrated in the lower right part of the figure. The upper figure depicts schematically the spins interacting with the lattice.}
\end{figure}  

To pursue our analysis, we assume a pair of magnon modes in a bipartite antiferromagnetic (AFM) material subjected to a vibrational mode in an off-resonant rotating magnetic field as shown in Fig.~ \ref{fig:illus}. The system Hamiltonian reads 
\begin{eqnarray}
\hat{\tilde{H} }  = \hat{\tilde{H} }_{m} + \hat{\tilde{H} }_{\text{m-m}}+ \hat{\tilde{H} }_\text{z}+\hat{\tilde{H} }_{\text{ph}}+ \hat{\tilde{H} }_{\text{m-ph}}
\label{Tol_H}
\end{eqnarray}
with each term defined as follows.
$\hat{\tilde{H} }_{\text{m}}$ and $\hat{\tilde{H} }_{\text{m-m}}$ correspond to single- and two-mode magnon Hamiltonian, respectively, at a given ${\bf k}$-vector. To find the explicit form of these terms, we consider an AFM spin Hamiltonian, $\hat{\tilde{H} }_s=\sum_{<i,j>}\hat{\mathbf{S}}_{i}\mathbb{I}_{ij}\hat{\mathbf{S}}_{j}$ with nearest neighbor interactions, where $\hat{\mathbf{S}}_{i}$ is the spin vector operator at lattice site $i$,  and $\mathbb{I}_{ij}$ is the bilinear interaction tensor matrix between sites $i$ and $j$, which involves all different types of interactions, e.g., Heisenberg interaction, Dzyaloshinskii-Moriya interaction, anisotropy
\cite{Azimi-mousolou2020,Azimi-mousolou2021,Azimi-mousolou2023,Shiranzaei2023}. By using the linearized Holstein-Primakoff transformation
\begin{eqnarray}
\text{Sublattice} \ A: \left\{
\begin{array}{lr}
\hat{S}^{z}_{i} = S-\hat{a}^{\dagger}_{i}\hat{a}_{i} , 
\nonumber\\
\hat{S}^{+}_{i}=\hat{S}^{x}_{i}+i\hat{S}^{y}_{i} \approx\sqrt{2S} \hat{a}_{i} , 
\end{array} \right.\nonumber\\
\text{Sublattice} \ B:\left\{
\begin{array}{lr}
\hat{S}^{z}_{j} = \hat{b}^{\dagger}_{j} \hat{b}_{j} - S , 
\nonumber\\
\hat{S}^{+}_{j}=\hat{S}^{x}_{j}+i\hat{S}^{y}_{j} \approx \sqrt{2S}\hat{b}^{\dagger}_{j}, 
\end{array}
\right. 
\end{eqnarray}
and Fourier transformation \cite{yuan2022}, one obtains $\hat{\tilde{H} }_s=\hat{\tilde{H} }_m + \hat{\tilde{H} }_{\text{m-m}}$ with
\begin{eqnarray}
\hat{\tilde{H} }_m&=&\hbar \omega_{a_{\bf k}} \hat{a}_{\bf k}^{\dagger}\hat{a}_{\bf k} + \hbar \omega_{b_{-{\bf k}}} \hat{b}_{-{\bf k}}^{\dagger}\hat{b}_{-{\bf k}}\nonumber\\
\hat{\tilde{H} }_{\text{m-m}}&=&\hbar g_{a_{\bf k}b_{-{\bf k}}}^* \hat{a}_{\bf k}^{\dagger}\hat{b}_{-{\bf k}}^{\dagger} + \hbar g_{a_{\bf k}b_{-{\bf k}}} \hat{a}_{\bf k} \hat{b}_{-{\bf k}}, 
\end{eqnarray}
for a given momentum ${\bf k}$-vector. Here, the bosonic annihilation operators $\hat{a}_{\bf k}$ and $\hat{b}_{-{\bf k}}$ represent two counter-propagating magnon modes that describe identifiable magnon modes 
associated with each sublattice of the AFM.

We further assume that the prepared magnon system is subjected to an off-resonant rotating magnetic field with frequency $\omega_0<\omega_{a_{\bf k}},\ \omega_{b_{-{\bf k}}}$.
An in-plane rotating magnetic field 
\begin{eqnarray}
{\bf B}=B_0(\cos(\phi_0+\omega_0t),\ \sin(\phi_0+\omega_0t), 0)
\end{eqnarray}
interacts with a spin system through the Zeeman term $\hat{\tilde{H} }_\text{z}=\sum_{i}\mu {\bf B}\cdot\mathbf{S}_{i}$. By using the same transformations as for $\hat{\tilde{H} }_s$, we obtain 
\begin{eqnarray}
\hat{\tilde{H} }_\text{z} &=& \hbar\left[\frac{\mu B_0\Gamma_{\bf k}}{\hbar}\right]\sqrt{\frac{SN}{2}}\left(e^{-i(\phi_0+\zeta_{\bf k}+\omega_0 t)}\hat{a}_{\bf k}+e^{i(\phi_0+\zeta_{\bf k}+\omega_0 t)}\hat{a}^{\dagger}_{\bf k}\right)\nonumber\\
&&+\hbar\left[\frac{\mu B_0\Gamma_{\bf k}}{\hbar}\right]\sqrt{\frac{SN}{2}}\left(e^{i(\phi_0+\zeta_{\bf k}+\omega_0 t)}\hat{b}_{-{\bf k}}+e^{-i(\phi_0+\zeta_{\bf k}+\omega_0 t)}\hat{b}^{\dagger}_{-{\bf k}}\right),\nonumber\\
\end{eqnarray}
for a given ${\bf k}$-vector. $B_0$ is the strength of the magnetic field, $N$ is the number of spins in the system, $S$ is the total spin per site, and $\Gamma_{\bf k} e^{-i\zeta_{\bf k}}=\frac{2}{N}\sum_{i\in A (B)}e^{-i{\bf k}\cdot {\bf r}_i}$ with ${\bf r}_i$ being the position vector of site $i$ in the corresponding sublattice. $\Gamma_{\bf k}$ and $\zeta_{\bf k}$ come about by 
Fourier transforming the Hamiltonian from real space to ${\bf k}$-space. It is assumed that the values of $\Gamma_{\bf k}$ and $\zeta_{\bf k}$ for sublattice $A$ are identical to those for sublattice $B$.

For the magnon-phonon interaction, we consider magnetoelastic coupling, where magnetic degrees of freedom interact with elastic displacements of atoms from their equilibrium positions.
Assuming single phonon process \cite{Upadhyaya1963} with translational and rotational symmetries, the quantized phonon and magnetoelastic spin-phonon coupling terms of the Hamiltonian read \cite{Haakon2019} 
\begin{eqnarray}
\hat{\tilde{H} }'_\text{ph} &=&\hbar\sum_{{\bf k}'}\omega_{c_{{\bf k}'}} \hat{c}_{{\bf k}'}^{\dagger}\hat{c}_{{\bf k}'},\nonumber\\
\hat{\tilde{H} }'_\text{m-ph} &=&\hbar\sum_{{\bf k}'}\sum_{{i,\ {\bf \delta}}}\sum_{\alpha,\ \beta}G^{\alpha\beta}_{{\bf k}'{\bf \delta}}\mathbf{S}^{\alpha}_{i}\mathbf{S}^{\beta}_{i+{\bf \delta}}\left(\hat{c}_{{\bf k}'}^{\dagger}+\hat{c}_{-{\bf k}'}\right)e^{i{\bf k}'\cdot r_{i}}, \nonumber\\
\end{eqnarray}
where $i$ is summed over all magnetic lattice sites and ${\bf \delta}$ denotes a vector pointing from lattice site $i$ to a neighbouring site. $G^{\alpha\beta}_{{\bf k}'{\bf \delta}}$ is the coupling strength between the spins in spatial directions $\alpha,\ \beta\in\{x,\ y,\ z\}$ and the elastic displacement mode at a given ${\bf k}'$. 
The operators $\hat{c}_{{\bf k}'}$ ($\hat{c}_{{\bf k}'}^{\dagger}$) represent annihilation (creation) of an elastic phonon mode with frequency $\omega_{c_{{\bf k}'}}$ and crystal momentum ${\bf k}'$.
Taking into account the magnetoelastic coupling only in the transverse direction, the interacting spin-phonon coupling term of the Hamiltonian reduces to 
\begin{eqnarray}
\hat{\tilde{H} }'_\text{m-ph}=\hbar\sum_{{\bf k}'}\sum_{{i,\ {\bf \delta}}}G^{zz}_{{\bf k}'{\bf \delta}}\mathbf{S}^{z}_{i}\mathbf{S}^{z}_{i+{\bf \delta}}\left(\hat{c}_{{\bf k}'}^{\dagger}+\hat{c}_{-{\bf k}'}\right)e^{i{\bf k}'\cdot r_{i}}.
\end{eqnarray}
By applying the same transformations to the spin degrees of freedom as for the other terms in the Hamiltonian above, the resulting magnetoelastic Hamiltonian takes the form
\begin{eqnarray}
\hat{\tilde{H} }'_\text{m-ph} &=&\hbar\sum_{{\bf k},\ {\bf k}'}\left[g^{({\bf k},\ {\bf k}')}_{ac}\hat{a}_{\bf k}^{\dagger}\hat{a}_{{\bf k}+{\bf k}'}  + g^{(-{\bf k},\ -
{\bf k}')}_{bc}\hat{b}_{-{\bf k}}^{\dagger}\hat{b}_{-{\bf k}-{\bf k}'}\right]\left(\frac{\hat{c}_{{\bf k}'}^{\dagger}+\hat{c}_{-{\bf k}'}}{\sqrt{2}}\right),\nonumber\\
\end{eqnarray}
where a term corresponding to a shift of each phonon mode is neglected.
This type of magnon-phonon interaction naturally describes the scattering of magnons with emission or absorption of phonons based on momentum conservation \cite{ Dany2019, streib2019, Zhang2016, woods2001, ruckriegel2014}.  
However, if a given pair of magnon modes $\hat{a}_{\bf k}$ and $\hat{b}_{-{\bf k}}$ are initially prepared in an excited state having large occupation number, the subsequent evolution will be dominated by interactions involving  $\hat{a}^{\dagger}_{\mathbf{k}} \hat{a}_{\mathbf{k}}$ and $\hat{b}^{\dagger}_{-\mathbf{k}} \hat{b}_{-\mathbf{k} }$, thereby forcing the phonon they couple with to have zero momentum, i.e., $\mathbf{k}'=0$. 
In the light of the dominant term, the effective magnon-phonon interaction Hamiltonian takes the form 
\begin{eqnarray}
\hat{\tilde{H} }_\text{m-ph} &=&\hbar\left[g^{({\bf k},\ 0)}_{ac}\hat{a}_{\bf k}^{\dagger}\hat{a}_{{\bf k}} + g^{(-{\bf k},\ 0)}_{bc}\hat{b}_{-{\bf k}}^{\dagger}\hat{b}_{-{\bf k}}\right]\left(\frac{\hat{c}_{0}^{\dagger}+\hat{c}_{0}}{\sqrt{2}}\right).
\nonumber\\
\label{m-ph-0k}
\end{eqnarray}
This implies that the phonon energy in the model system can be written as
\begin{eqnarray}
\hat{\tilde{H} }_\text{ph} =\hbar\omega_{c_{0}} \hat{c}_{0}^{\dagger}\hat{c}_{0},
\end{eqnarray}
which physically corresponds to zone-center optical phonon modes.
Note that phonon contributions from non-zero $\mathbf{k}'$ are included implicitly in the dynamics of the system as dissipation and noise terms in the quantum Langevin equations (QLEs) below.
It is worth noting that the type of bosonic Hamiltonian expressed in Eq.\eqref{m-ph-0k} is commonly used in optomechanical systems \cite{browne2017, sanavio2020}.   

By collecting the above terms for a system with magnons of a specific value of $\bf k$ being prepared, the relevant Hamiltonian in the rotating frame, given by applying
$\hat{U}(t)=e^{i\omega_0 t[\hat{b}^{\dagger}_{-\mathbf{k}} \hat{b}_{-\mathbf{k}}-\hat{a}^{\dagger}_{\mathbf{k}} 
\hat{a}_{\mathbf{k}}]}$, takes the following 
form
\begin{eqnarray}
    \hat{H}&=&i\hbar \frac{d\hat{U}}{dt}\hat{U}^{\dagger}+\hat{U}\hat{\tilde{H} }\hat{U}^{\dagger}\nonumber\\
    &=&\hbar \left(\omega_{a_{\bf k}}+\omega_0\right) \hat{a}_{\bf k}^{\dagger}\hat{a}_{\bf k} + \hbar \left(\omega_{b_{-{\bf k}}}-\omega_0\right) \hat{b}_{-{\bf k}}^{\dagger}\hat{b}_{-{\bf k}}\nonumber\\
    && + \hbar g_{a_{\bf k}b_{-{\bf k}}}^* \hat{a}_{\bf k}^{\dagger}\hat{b}_{-{\bf k}}^{\dagger} + \hbar g_{a_{\bf k}b_{-{\bf k}}} \hat{a}_{\bf k} \hat{b}_{-{\bf k}}
    \nonumber\\
    && + \hbar\Omega\left(e^{-i(\phi_0+\zeta_{\bf k})}\hat{a}_{\bf k}+e^{i(\phi_0+\zeta_{\bf k})}\hat{a}^{\dagger}_{\bf k}\right)\nonumber\\
    &&+ \hbar\Omega\left(e^{i(\phi_0+\zeta_{\bf k})}\hat{b}_{-{\bf k}}+e^{-i(\phi_0+\zeta_{\bf k})}\hat{b}^{\dagger}{-{\bf k}}\right)\nonumber\\
    && + \hbar\omega_{c_0}\hat{c}_{0}^{\dagger}\hat{c}_{0}+ \hbar\left[g_{a_{\mathbf{k}}c_0}\hat{a}^{\dagger}_{\mathbf{k}} \hat{a}_{\mathbf{k}}+g_{b_{-\mathbf{k}}c_0}\hat{b}^{\dagger}_{-\mathbf{k}} \hat{b}_{-\mathbf{k}}\right] \left( \frac{\hat{c}^{\dagger}_{ 0 } + \hat{c}_{0}}{\sqrt{2}} \right)\nonumber\\
    \label{eq:Htot_raw}
\end{eqnarray}
with $\Omega=\left[ \frac{\mu B_0 \Gamma_{\bf k}}{\hbar}\right] \sqrt{\frac{SN}{2}}$. Note that the parameters $\Omega$ and $\zeta_{\bf k}$ can be tuned by the strength $B_0$ and azimuthal angle $\phi_0$ of the off-resonant rotating magnetic field. Thus, without loss of generality, we assume $\phi_0+\zeta_{\bf k}\equiv 0$.

To summarize, the model Hamiltonian takes the form
\begin{eqnarray}
\hat{H} & = & \hbar \omega_a \hat{a}^{\dagger}\hat{a} + \hbar \omega_b \hat{b}^{\dagger}\hat{b} + \hbar \omega_c \hat{c}^{\dagger}\hat{c}\nonumber \\
&&+\hbar \Omega (\hat{a}+\hat{a}^{\dagger}) + \hbar \Omega(\hat{b}+\hat{b}^{\dagger})
\nonumber \\
&&+ \hbar g_{ab}^* \hat{a}^{\dagger}\hat{b}^{\dagger} + \hbar g_{ab} \hat{a} \hat{b}
\nonumber \\
&&+ \hbar \left[g_{ac} \hat{a}^{\dagger}\hat{a} + g_{bc} \hat{b}^{\dagger}\hat{b}\right]\left(\frac{\hat{c}+\hat{c}^{\dagger}}{\sqrt{2}}\right),
\label{Mod_H}
\end{eqnarray}
where we drop the subscripts $\mathbf{k}, -\mathbf{k}, 0$ for simplicity. We assume that the magnon modes in the AFM have the same frequency $\omega$ that are split by an off-resonant external rotating magnetic field with frequency $\omega_0$. Thus, we have the detuned frequencies $\omega_a = \omega + \omega_0$ and $\omega_b = \omega - \omega_0$ of the two magnon modes. The magnon-magnon coupling with strength $g_{ab}$ is known to give rise to magnon squeezing and entanglement in each magnon eigenstate in AFMs. \cite{Azimi-mousolou2020,Azimi-mousolou2021,Azimi-mousolou2023,Shiranzaei2023}.

Physical systems that are relevant for this investigation should have zone center (optical) phonons that couple to magnons with finite crystal momentum (up to the zone boundary). Typically this happens when the frequencies are of similar magnitude. There are many reports in the literature on such materials \cite{Haakon2019, Sheng2021, streib2019}.
Thus, we pay a particular attention to  the frequency ratio $\omega_c/\omega = 1$, although, as Figs.~\ref{fig:ratios} and \ref{fig:couplings} below show, our results are valid for a wider range of $\omega_c/\omega$ as well.

\section{Dynamics of the system}
We consider the system under dissipation caused by thermal fluctuations, uncontrolled coupling to 
other modes (e.g., phonons with non-zero momentum $\mathbf{k}'$), and Brownian motion. Such dissipative dynamics can be described by nonlinear QLEs \cite{Vittorio2001}.

To specify the dynamics of the system, it is handy to introduce dimensionless quadratures
 \begin{eqnarray}
 & \hat{X} = (\hat{a} + \hat{a}^{\dagger} ) / \sqrt{2}, \ \ \ \  \hat{Y} = i (\hat{a}^{\dagger} - \hat{a}) / \sqrt{2}, \nonumber \\ 
 & \hat{x} = (\hat{b} + \hat{b}^{\dagger} ) / \sqrt{2}, \ \ \ \ \hat{y} = i (\hat{b}^{\dagger} - \hat{b}) / \sqrt{2},     
 \end{eqnarray}
for the magnon modes $\hat{a} , \hat{b}$. The dissipative dynamics for magnon modes are described by
\begin{eqnarray}
    \dot{\hat{O}} = \frac{i}{\hbar} [\hat{H}, \hat{O}]-\kappa_o \hat{O}+\sqrt{2\kappa_o}\hat{O}^{in},
    \label{EQmotion1}
\end{eqnarray} 
where $\kappa_o$ is the dissipation rate, $\hat{O}^{in}=(\hat{o}^{in} + \hat{o}^{in\dagger}) / \sqrt{2}$ or $\hat{O}^{in}=i(\hat{o}^{in\dagger}-\hat{o}^{in}) / \sqrt{2}$ with
$\hat{o}=\hat{a} , \hat{b}$, is the input noise operator associated with the magnon modes $\hat{O}=\hat{X},\hat{x}$ or $\hat{O}=\hat{Y},\hat{y}$, respectively. The input noise operators are characterized by the correlation functions
\begin{eqnarray}
& \langle  \hat{o}^{in\dagger}(t) \hat{o}^{in}(t') \rangle =2 \kappa_o N(\omega_o) \delta(t-t'), \nonumber \\
& \langle  \hat{o}^{in}(t) \hat{o}^{in\dagger}(t') \rangle =2 \kappa_o [N(\omega_o) +1] \delta(t-t'),
\end{eqnarray}
 with the equilibrium thermal mean magnon occupation numbers $N (\omega_o)= [\exp(\hbar \omega_o / k_B T) - 1]^{-1}$. 

In a similar manner we may define 
\begin{eqnarray}
     \hat{q} = (\hat{c} + \hat{c}^{\dagger} ) \sqrt{2},  \ \ \ \ 
    \hat{p} = i (\hat{c}^{\dagger} - \hat{c}) / \sqrt{2},
\end{eqnarray}
to be the dimensionless quadratures associated with the mechanical phonon mode. A Markovian description of quantum Brownian motion for a mechanical mode with large quality factor $Q=\omega_c/\gamma_c\gg 1$ is set by \cite{Jie2018, Vittorio2001, vitali2007}
\begin{eqnarray}
\dot{\hat{q}} = \frac{i}{\hbar} [\hat{H}, \hat{q}], \ \  \dot{\hat{p}} = \frac{i}{\hbar} [\hat{H}, \hat{p}]-\gamma_c \hat{p}+\hat{\xi}, \label{EQmotion2}
\end{eqnarray}
where the mechanical dissipation rate $\gamma_c$ is associated mainly with phonons of non-zero momentum $\mathbf{k}'$, which are not considered in the Hamiltonian $\hat{H}$. The input noise operator $\hat{\xi}$ is given by the correlation function 
\begin{eqnarray}
\langle  \hat{\xi}(t) \hat{\xi}(t') + \hat{\xi}(t') \hat{\xi}(t)\rangle \simeq 2\gamma_c [2 N_c(\omega_c) +1] \delta(t-t'),
\end{eqnarray}
where $N_c(\omega_c)=[\exp(\hbar \omega_c / k_B T) - 1 ]^{-1}$ is the equilibrium thermal mean phonon occupation number. 

Therefore, the corresponding nonlinear QLEs, in terms of dimensionless quadratures, take the form
\begin{eqnarray}
         \dot{\hat{X}} & = & -\kappa_a \hat{X} + \omega_a \hat{Y} - {\rm Im}(g_{ab}) \hat{x} - {\rm Re}(g_{ab}) \hat{y} + g_{ac} \hat{Y} \hat{q} 
         \nonumber \\ 
         & & + \sqrt{2}\Omega + \hat{X}^{in}, \nonumber \\ 
         \dot{\hat{Y}} & = & -\omega_a \hat{X} - \kappa_a \hat{Y}  - {\rm Re}(g_{ab}) \hat{x} + {\rm Im}(g_{ab}) \hat{y} - g_{ac} \hat{X}\hat{q} +  \hat{Y}^{in \dagger} , \nonumber \\
         \dot{\hat{x}} & = &  - {\rm Im}(g_{ab}) \hat{X} - {\rm Re}(g_{ab}) \hat{Y} -\kappa_b \hat{x} + \omega_b \hat{y}   + g_{bc} \hat{y} \hat{q} 
         \nonumber \\ 
         & & + \sqrt{2}\Omega +  \hat{x}^{in}, \nonumber \\ 
         \dot{\hat{y}} & = &  - {\rm Re}(g_{ab}) \hat{X}  + {\rm Im}(g_{ab}) \hat{Y} -\omega_b \hat{x} - \kappa_a \hat{y}  - g_{bc} \hat{x}\hat{q} +  \hat{y}^{in \dagger}, \nonumber \\
         \dot{\hat{q}} & = & \omega_c \hat{p}, \nonumber \\
         \dot{\hat{p}} & = &  - \frac{g_{ac}}{2} (\hat{X}^{2}+\hat{Y}^{2}-1) - \frac{g_{ac}}{2} (\hat{x}^{2}+\hat{y}^{2}-1) 
          -\omega_c \hat{q} -\gamma_c \hat{p} + \hat{\xi}.\nonumber\\
\label{QLEs}
\end{eqnarray}

\section{Results and discussion}

We now show that it is possible to maintain magnon-magnon entanglement in an equilibrium setting. We focus on the steady-state regime, where $ \frac{d}{dt} \langle \hat{O} \rangle = 0$ for $\hat{O} = \hat{X}, \ \hat{Y}, \hat{x}, \ \hat{y}, \hat{p}, \hat{q}$. Any operator can then be written as a steady-state expectation value plus an additional quantum fluctuation term, i.e., $\hat{O} = \langle \hat{O} \rangle + \delta \hat{O}(t)$ \cite{Mancini1994}. Imposing the steady-state conditions on Eq.~(\ref{QLEs}) and retaining fluctuations up to linear order, we obtain a set of linearized QLEs
\begin{eqnarray}
 \dot{u}(t) = A u(t) + n(t), 
 \label{linearQLEs}
\end{eqnarray}
where $ u(t) = (\delta \hat{X}, \delta \hat{Y}, \delta \hat{x}, \delta \hat{y}, \delta \hat{q}, \delta \hat{p})^{\rm T}$ and $ n(t) = (\hat{X}^{in} , \hat{Y}^{in}, \hat{x}^{in}, \hat{y}^{in}, 0, \hat{\xi} )^{\rm T}$. Information about steady-state expectation values and coupling constants is encoded into the drift matrix:
\begin{eqnarray}
A = \begin{bmatrix}
            -\kappa_a     & \Delta_{ac}    & -{\rm Im} (g_{ab})    & -{\rm Re}(g_{ab})     &  M_Y   &  0 \\
            -\Delta_{ac}  & -\kappa_a      & -{\rm Re} (g_{ab})    & {\rm Im} (g_{ab})      &  -M_X  &  0 \\
            -{\rm Im} (g_{ab})     & -{\rm Re} (g_{ab})      & -\kappa_b    & \Delta_{bc}   &  M_y   &  0 \\
            -{\rm Re}(g_{ab})     &  {\rm Im}(g_{ab})      & -\Delta_{bc} &  -\kappa_b    &  -M_x  &  0 \\ 
                0         &      0         &    0         &    0          &     0               & \omega_p \\ 
              -M_X   &   -M_Y    &   -M_x  &   -M_y   &  -\omega_p   & -\gamma_c \\
        \end{bmatrix}.\nonumber \\
\label{DM}
\end{eqnarray}
Here, $\Delta_{ac} = \omega_a + g_{ac}\langle \hat{q}\rangle$ and $\Delta_{bc} = \omega_b + g_{bc}\langle \hat{q}\rangle$ are the effective magnon frequency detunings induced by the magnon-phonon interaction. The effective magnon-phonon couplings are given by $ M_X = g_{ac}\langle \hat{X}\rangle, \ M_Y = g_{ac}\langle \hat{Y}\rangle $, and  $ M_x = g_{bc}\langle \hat{x}\rangle, \ M_y = g_{bc}\langle \hat{y}\rangle $. Assuming low dissipation of the magnon modes compared to their frequencies in typical AFM structures \cite{Kruglyak2010, Lenk2011}, i.e., $\kappa_a, \ \kappa_b \ll \Delta_{ac}, \Delta_{bc} $, we obtain the steady-state solutions as 
\begin{eqnarray}
\langle X \rangle & = & \frac{\sqrt{2} \Omega \rm Im(g_{ab})}{\Delta_{ac} \Delta_{bc} - \mid g_{ab} \mid^2} ,\  \langle Y \rangle =\frac{ -\sqrt{2} \Omega \left [ \Delta_{bc} + \rm Re(g_{ab}) \right ]}{\Delta_{ac} \Delta_{bc} - \mid g_{ab} \mid^2 },
\nonumber \\ 
\langle x \rangle & = & \frac{\sqrt{2} \Omega \rm Im(g_{ab})}{\Delta_{ac} \Delta_{bc} - \mid g_{ab} \mid^2} ,\  \langle y \rangle =\frac{ -\sqrt{2} \Omega \left [ \Delta_{ac} + \rm Re(g_{ab}) \right ]}{\Delta_{ac} \Delta_{bc} - \mid g_{ab} \mid^2 },
\nonumber \\ 
 & &\frac{\omega_p}{\Omega^2} \langle q \rangle + \frac{g_{ac} (\Delta_{bc} + g_{ab} )^2 + g_{bc} (\Delta_{ac} + g_{ab})^2}{ (\Delta_{ac} \Delta_{bc} - \mid g_{ab} \mid^2) ^2 } = 0.
\label{SimpPolyCase3}
\end{eqnarray}

Once the frequencies and exchange parameters are set as parameters of the model Hamiltonian in Eq. \eqref{Mod_H}, the explicit form of the drift matrix $A$ mainly depends on the steady-state value of $\langle \hat{q}\rangle$, which is given as a root of the polynomial of degree five specified by the last algebraic expression in Eq.~\eqref{SimpPolyCase3}. 
As a frequency shift in magnon modes, only the real-valued roots of $\langle \hat{q}\rangle$ are physically meaningful. For a given set of parameter values, the polynomial has only one real-valued root. 

The linear form of QLEs in Eq.\ \eqref{linearQLEs} ensures that the dynamics of quantum fluctuations is Gaussian and can be completely characterized by the corresponding covariance correlation matrix (CCM) \cite{Adesso2004}. By using the fact that different components of the input noise vector $n(t)$ are uncorrelated, the steady-state CCM denoted by $V$ can be obtained through the Lyapunov equation \cite{Jie2018, vitali2007}
\begin{eqnarray}
     A\cdot V + V \cdot A^T = -D, 
 \label{eq:Lyapunov}    
\end{eqnarray}
where $ D = \text{diag}[ \kappa_a (2N_a + 1), \kappa_a (2N_a + 1),\kappa_b (2N_b + 1),\kappa_b (2N_b + 1), 0,\gamma_c (2N_p + 1)]$ is the diffusion matrix. Solving Eq.~\eqref{eq:Lyapunov} for $V$, one can evaluate the bipartite magnon-magnon entanglement by computing the logarithmic negativity of two-mode Gaussian states defined as \cite{Jie2018, Vidal2002, Adesso2004, Plenio2005}
\begin{eqnarray}
     E_N \equiv \max[ 0, -\ln (2 \tilde{\nu}_{-})],
\label{eq:LN}
\end{eqnarray}
where $\tilde{\nu}_-$ is the minimum symplectic eigenvalue of a matrix, $ V_4$, being the $ 4 \times 4$ reduced CCM obtained by projecting $V$ onto the two magnon modes. This eigenvalue can be computed as $\tilde{\nu}_- = \min \ {\rm eig} \  |i\Upsilon \tilde{V}_4| $, for the symplectic matrix $\Upsilon = \oplus_{j=1}^2 i \sigma_y $ (with $ \sigma_y $ being the Pauli-$y$ matrix) and the partial transpose $\tilde{V}_4 = P_{1|2} V_4 P_{1|2}$ with $ P_{1|2} = \text{diag}( 1, -1, 1, 1) $. 

\begin{figure}[h]
\centering
\includegraphics[width=.7\textwidth]
{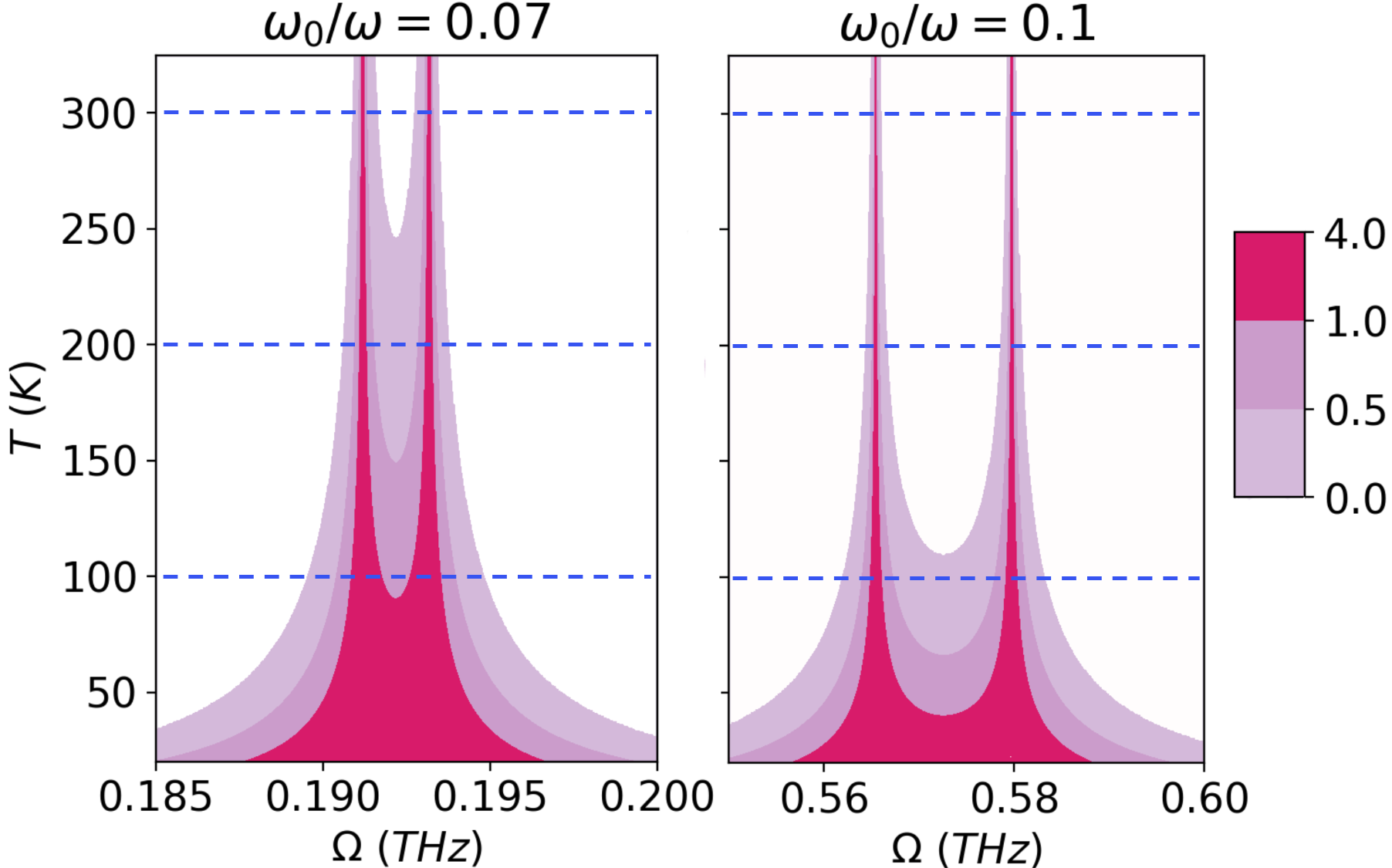}
\caption{(Color online) Phonon-driven magnon-magnon entanglement $E_N$ as a function of
temperature and the frequency $\Omega$. Note that $\Omega $ is associated with the strength of the off-resonant rotating magnetic field, $B_0$ (see discussion below Eq.\eqref{eq:Htot_raw} ).  The color bar displays a qualitative illustration of the entanglement, where the white area indicates the separable zone while the colored zones shows non-zero magnon-magnon entanglement $E_N$ in the $\Omega-T$ plane. Double peaks are associated with the degeneracy of two polarized magnons in AFM systems, which is broken in the presence of non-zero magnetic field frequency $\omega_0$. The gap between the peaks is proportional to the frequency difference of the two magnons. The ratio $\omega_0/\omega$ with respect to the magnon frequency at the degeneracy point, i.e., $\omega$, allows one to control the magnon-magnon entanglement region in the $\Omega-T$ parameter space. The lower the ratio $\omega_0/\omega$, the weaker the external magnetic field required to achieve high entanglement at a given temperature (see also Fig.~ \ref{fig:Bw0}). Here, the remaining parameters of the Hamiltonian in Eq.~(\ref{Mod_H}) satisfy $\omega_c/\omega=1$ and $g_{ac}/g_{ab}=g_{bc}/g_{ab}=0.001$.}
\label{fig:BT}
\end{figure}  

\begin{figure}[h]
\centering
\includegraphics[width=.65\textwidth]{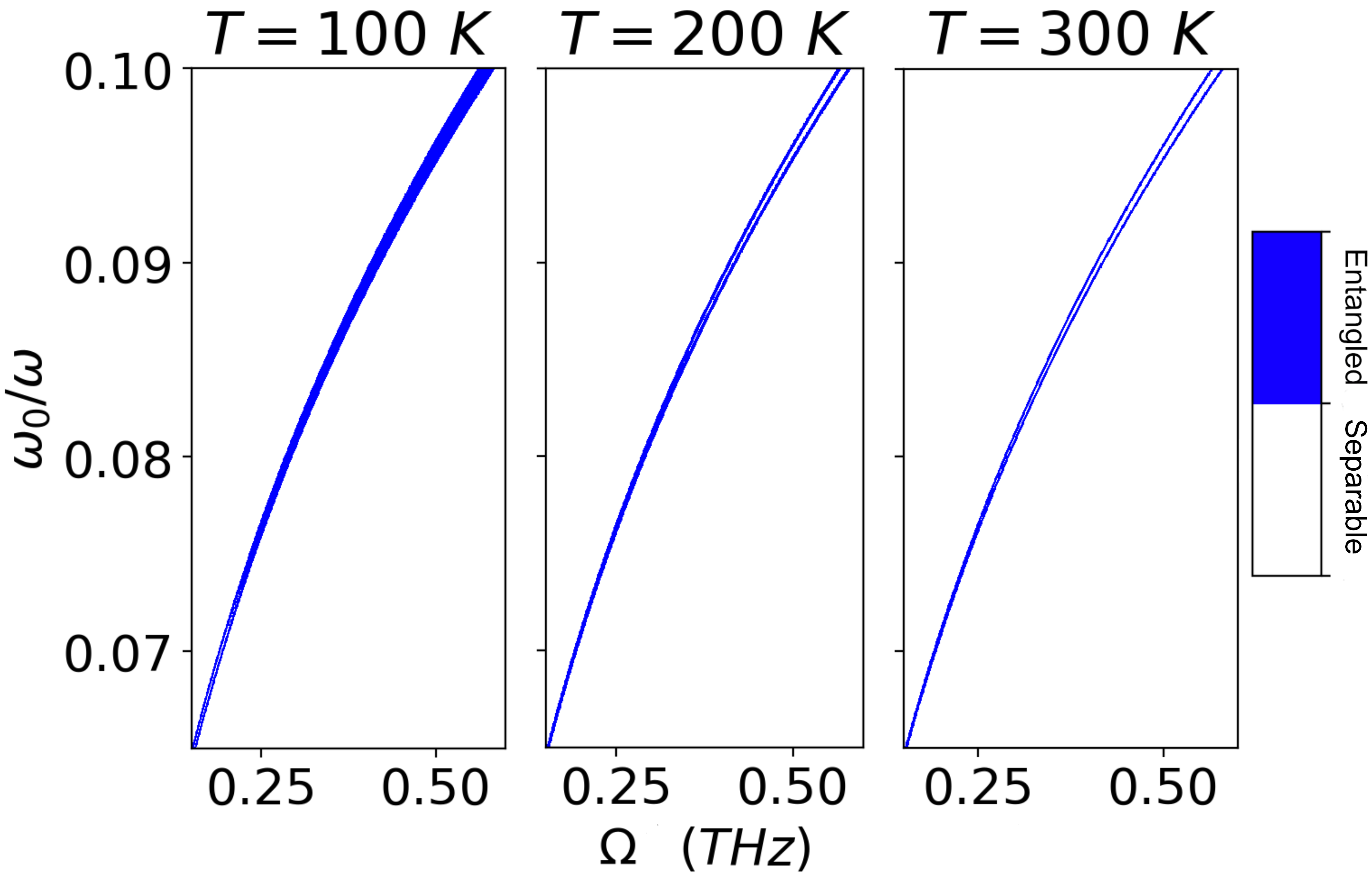} 
\caption{(Color online) Magnon-magnon entanglement $E_N$ as a function of the frequency ratio $\omega_0/\omega$ and external magnetic field strength $\Omega \propto B_0$ at three temperatures $T = 100$ K (left), $T = 200$ K (middle) and $T= 300$ K (right). The dark blue color specifies non-zero entanglement (entangled), all the other uncolored regions depict zero-entanglement (separable). We keep $g_{ac}/g_{ab}=g_{bc}/g_{ab} = 0.001$ and $\omega_c / \omega = 1$, which are the same as the parameter values used for Fig.~ \ref{fig:BT}.}
\label{fig:Bw0}
\end{figure}

\begin{figure}[h]
\centering
\includegraphics[width=.65\textwidth]{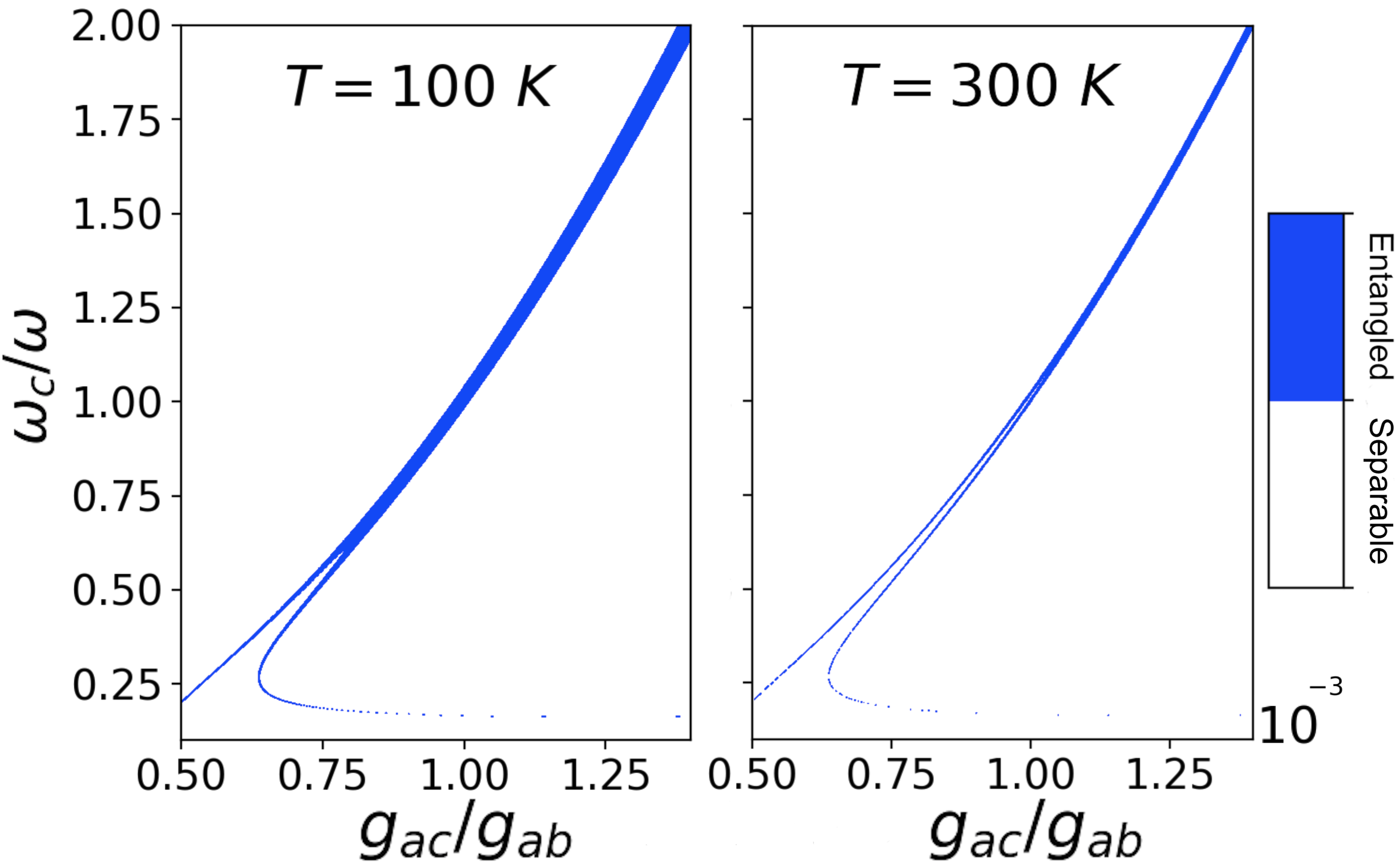} 
\caption{(Color online)
Segments of magnon-magnon entanglement $E_N$ in Fig.~ \ref{fig:BT} are shown by the effect of two factors, the magnon-phonon coupling ratios, $g_{ac}/g_{ab}$ ($=g_{bc}/g_{ab}$), and the phonon-magnon frequency ratio 
$\omega_{c}/\omega$.
The blue area marks a range in the $g_{ac}/g_{ab}-\omega_{c}/\omega$ parameter domain, which gives the same magnon-magnon entanglement pattern as in Fig.~ \ref{fig:BT}, all the other uncolored regions depict zero-entanglement (separable). Although the plots correspond to the non-zero magnon-magnon entanglement at $\Omega = 0.1932$ THz,  $\omega_0/\omega = 0.07$, as well as temperatures $T = 100$ K (left) and $T = 300$ K (right), the same range of phonon-magnon frequency ratio and magnon-phonon coupling ratio is obtained for other points of the entanglement diagram in Fig.~ \ref{fig:BT}. The higher the temperature the narrower the range of relative parameters.}
\label{fig:ratios} 
\end{figure}

\begin{figure}[h]
    \centering
    \includegraphics[width=.65\textwidth]{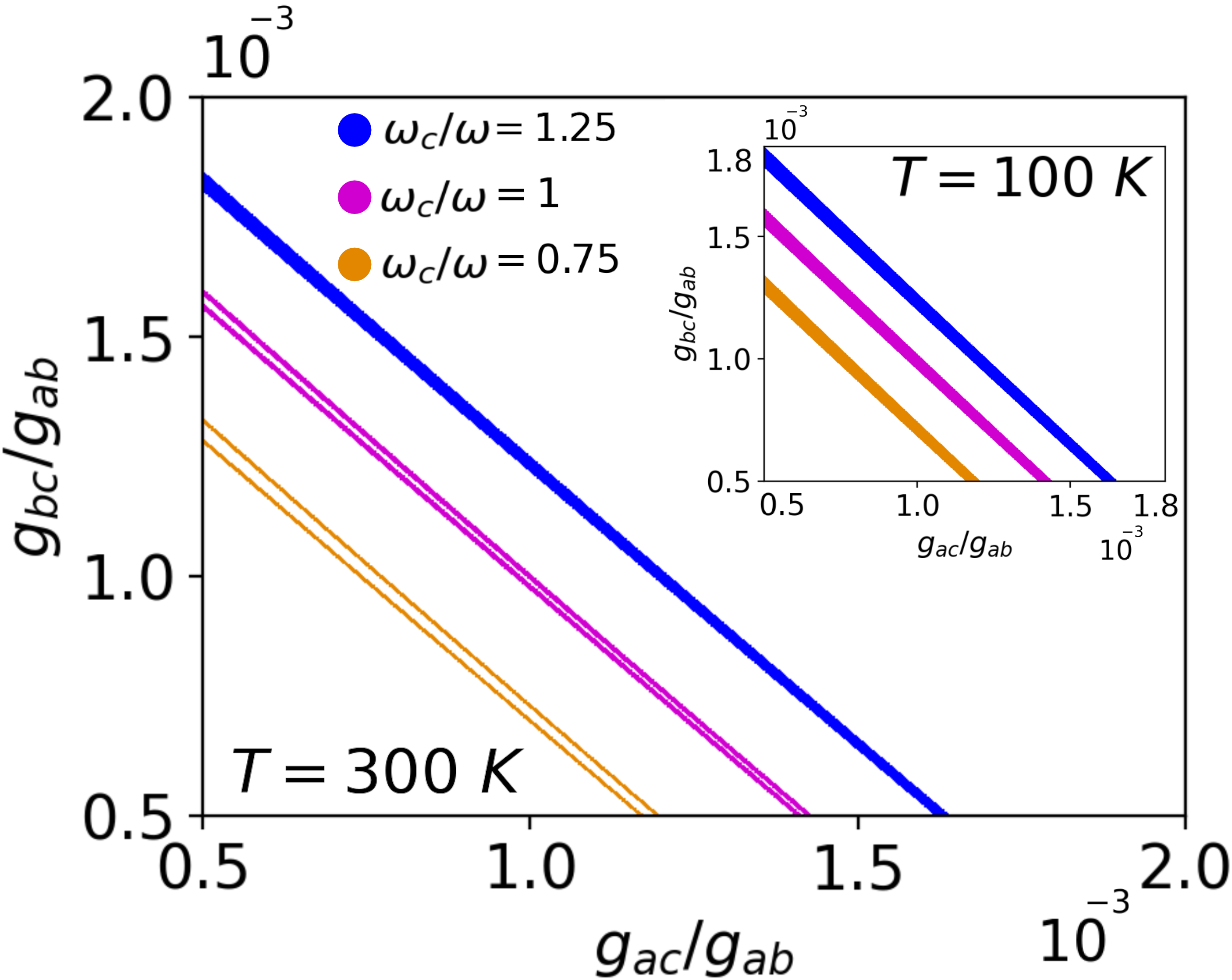}
    \caption{(Color online) The range of asymmetric magnon-phonon couplings, which allows for the same magnon-magnon entanglement pattern in Fig.~ \ref{fig:BT}, for a given phonon frequency. The contour plots correspond to the non-zero magnon-magnon entanglement segment at $\Omega = 0.1932$ THz,  $\omega_0/\omega = 0.07$ and temperature $T = 300$ K ($T = 100$ K for inset) for $\omega_c/ \omega = 0.75$ (yellow), $\omega_c/ \omega = 1$ (pink) and $\omega_c/ \omega = 1.25$ (blue). The uncolored regions depict zero-entanglement (separable). Similar patterns occur for any other point in the entangling diagram of Fig.~ \ref{fig:BT} and any other value of $\omega_c/ \omega$. The higher the temperature the narrower the range of asymmetric magnon-phonon couplings for a given $\omega_c/\omega$.}
\label{fig:couplings} 
\end{figure}

The magnon-magnon logarithmic negativity in the steady-state limit as a function of temperature $T$ and external magnetic field strength $\Omega$  is shown in Fig.~\ref{fig:BT} for $\omega_c/\omega=1$. 
Note that in the following numerical analysis, $\gamma_c$ is taken to be $10^2 \ \rm{Hz}$, and $\kappa_{a, b} \approx 1 \ \rm{GHz}$.
As can be seen, our analysis reveals magnon-magnon entanglement at room temperature under the phonon mode effect. 
The high-temperature result for magnon-magnon entanglement is reproduced across a wide range of Hamiltonian parameters through complementary plots in Figs. \ref{fig:Bw0}, \ref{fig:ratios}, and \ref{fig:couplings}.
For any given rotation frequency $\omega_0$ of the magnetic field, there always exists a certain range of the field amplitude $\Omega$, where magnon-magnon entanglement can be realized at high temperature. Moreover, the entangled region including the high-temperature entanglement zone in the $\Omega-T$ parameter space can be efficiently placed around practically feasible parameter values by controlling the frequency ratio $\omega_0/\omega$. In other words, magnon-magnon entanglement can be realized for any positive value of $\Omega$ at a given temperature $T$, provided the frequency ratio $\omega_0/\omega$ is properly adjusted. 
The results depicted in Fig.~\ref{fig:BT} were qualitatively replicated across a range of magnon-phonon coupling
ratio $g_{ac}/g_{ab} = g_{bc}/g_{ab}\in [10^{-5}, 10^{-3}]$. This shows that, regardless of the strength of the magnon-phonon interaction in the reasonable range relative to the coupling strength between the two magnon modes, a similar outcome as presented in Fig.~ \ref{fig:BT} can be achieved. Thus, the strength of magnon-phonon interaction, as treated here, does not have a detrimental effect on the phase transition of the magnon-magnon entanglement. It is instead so that, in the steady-state limit, the phonon degrees of freedom have a key role in establishing magnon-magnon entanglement, in the sense that in the absence of phonons, the two magnon modes are completely disentangled. We demonstrate this by noting that the Hamiltonian in Eq.(\ref{Mod_H}) reduces to
\begin{eqnarray}
\hat{H}_{ab} & = & \hbar \omega_a \hat{a}^{\dagger}\hat{a} + \hbar \omega_b \hat{b}^{\dagger}\hat{b} \nonumber \\
&& +\hbar \Omega (\hat{a}+\hat{a}^{\dagger}) + \hbar \Omega(\hat{b}+\hat{b}^{\dagger})
\nonumber \\
&& + \hbar g_{ab}^* \hat{a}^{\dagger}\hat{b}^{\dagger} + \hbar g_{ab} \hat{a} \hat{b}, 
\label{eq:Mod_H_mm}  
\end{eqnarray}
when the phonon mode is absent. Through the similar method as shown above, we obtain the corresponding drift matrix
\begin{eqnarray}
    A_{ab} = \begin{bmatrix}
            -\kappa_a     & \omega_{a}    & -{\rm Im} (g_{ab})    & -{\rm Re}(g_{ab})  \\
            -\omega_{a}  & -\kappa_a      & -{\rm Re} (g_{ab})    & {\rm Im} (g_{ab})   \\
            -{\rm Im} (g_{ab})     & -{\rm Re} (g_{ab})      & -\kappa_b    & \omega_{b} \\
            -{\rm Re}(g_{ab})     &  {\rm Im}(g_{ab})      & -\omega_{b} &  -\kappa_b \\ 
        \end{bmatrix}.
\label{eq:A_ab}
\end{eqnarray}
and numerically confirm that $E_N^{ab} = 0$, for any choice of physical parameters of the spin Hamiltonian. 
We employ the drift matrix from Eq.\ \eqref{eq:A_ab} to solve Eq.\ \eqref{eq:Lyapunov} for the CCM, i.e., $V$. Subsequently, we determine the minimum eigenvalue $\nu_{-}$ for evaluating for evaluating the logarithmic negativity. The procedure is repeated for a wide range of distinct parameter values within the drift matrix described in Eq.~\eqref{eq:A_ab}. In all cases, $\tilde{\nu}_{-}$ consistently exceeds $\frac{1}{2}$ at any temperature, indicating a complete absence of entanglement. This indicates that the nontrivial entanglement diagram obtained in Fig.~\ref{fig:BT} is dissipation- and phonon-driven. 

To elaborate further on this point, in Fig.~\ref{fig:Bw0} we show how some cuts of the entangled region in Fig.~\ref{fig:BT} at three different temperatures move along the parameter domain of $\Omega$ upon varying the control parameter $\omega_0/\omega$. 
The plots indicate that for a given value of $\Omega$, there is a narrow interval for $\omega_0/\omega$ where magnon-magnon entanglement is non-zero and that this interval depends on temperature. The appearance of two lines in these plots at higher values of $\omega_0/\omega$ and temperatures corresponds to the two peaks in Fig.~\ref{fig:BT}. These peaks result from the detuning frequencies of the two magnon modes, where $\omega_a = \omega + \omega_0$ and $\omega_b = \omega - \omega_0$. As $\omega_0$ approaches zero, causing the two magnon modes to become degenerate, the two peaks converge and coincide at $\omega_0=0$, as indicated in Fig.~\ref{fig:Bw0}.

Figures \ref{fig:BT} and \ref{fig:Bw0} are plotted for fixed value of the frequency ratio $\omega_c/\omega=1$ and fixed ratios $g_{ac}/g_{ab}$ and $g_{bc}/g_{ab}$, and shows segments of entanglement in a parameter space of magnetic field strength and frequency ratio $\omega_0/\omega$. In a search for a real material that can serve as an optimal host for robustly entangled magnon modes, it is instructive to also understand the dependence of entanglement on these material parameters. This is shown in Fig. \ref{fig:ratios} and \ref{fig:couplings}. One can see that, in a wide range of couplings and phonon frequencies, it is possible to have non-zero entanglement at rather large (and even ambient) temperatures if frequency and amplitude of the rotating magnetic field are tuned to proper values. Fig.~ \ref{fig:ratios} demonstrates that for any given magnon-phonon coupling $g_{ac}=g_{bc}$, there always exists a narrow window of phonon frequencies, where the same magnon-magnon entanglement as shown in the Fig.~ \ref{fig:BT} occurs.    

We further examine the effect of asymmetric magnon-phonon couplings on the magnon-magnon entanglement region. For given $\omega_c/\omega$, Fig.~ \ref{fig:couplings} illustrates a wide range of different coupling strengths $g_{ac}\ne g_{bc}$, which allows the same entanglement phase diagram as in Fig.~ \ref{fig:BT}. Figure \ref{fig:couplings} shows that, even up to as high temperatures as $T=300$ K, for different values of coupling rates varying in a wide range, there is a value of phonon frequency $\omega_c/\omega$ that guarantees non-zero magnon-magnon entanglement. Besides, it implies that a broad family of antiferromagnetic materials are likely to have a similar magnon-magnon entanglement diagram, as shown in Fig.~\ref{fig:BT}.

There are numerous suitable classes of AFM compounds that can maintain the steady-state entanglement of magnon modes within the AFM structure, even up to room temperature. Among them, NiO is a concrete example of a crystalline AFM \cite{Roth1958, Moriyama2019, Tomlinson1955, Hiroyoshi2003} with high Néel temperature of $T_N^{{\rm NiO}} = 523K$ and large magnon lifetime. Notably, it has been shown that the zone center phonon frequency matches the magnon frequency at finite ${\bf k}$ frequencies $\sim$ 11.3 THz and $\sim$ 17.3 THz \cite{Haakon2019}.
The perovskite manganites also form a promising class of materials in this regard, with many compounds of varying chemical composition that have been synthesized (see, e.g.,  Refs.~\cite{Changting2021,Rini2007}). For these systems, upon chemical modulation, both the magnetic and lattice properties can be tuned. This has particular relevance for finding compositions that allow for finer adjustments of the magnon and phonon frequencies relevant for the results of the present investigation.

We end our discussion with a remark that the dissipative magnon entanglement studied here is of the mixed-state type and is achieved in the asymptotic steady-state limit. This is different from the entanglement observed in the pure energy eigenstates of ideal closed magnon systems studied previously in Ref.\ \cite{Azimi-mousolou2020}.

\section{Conclusion}
We have examined bipartite magnon-magnon entanglement in a general setting of an AFM material subjected to a dissipative elastic displacement (phonon) mode, thermal noise, and an external rotating magnetic field. We have provided evidence that it is possible to observe phonon-driven entanglement between the magnon modes of the two sublattices even at room temperature. 
The high-temperature magnon-magnon entanglement for a given material and magnon-phonon coupling, can be ensured by tuning the external magnetic field frequency and amplitude in a physically reasonable range of scales. The presence of a non-zero magnon-phonon coupling is necessary for maintaining entanglement between magnons in antiferromagnets in a dissipative steady-state regime, while the strength of the coupling does not significantly affect the existence of this entanglement.
It would be interesting to examine whether AFMs can be used as a resource for different quantum information applications, for instance quantum state transfer \cite{cirac97}, and to study the potential role of magnon-magnon entanglement in such processes.

The authors acknowledge financial support from the Knut and Alice Wallenberg Foundation through Grant No. 2018.0060.  A.B. and O.E. acknowledges eSSENCE.
A.D.~acknowledges financial support from the Swedish Research Council (VR) through Grants No.~2019-05304 and 2016-05980. 
O.E.~acknowledges support by the Swedish Research Council (VR), the Foundation for Strategic Research (SSF), the European Research Council (854843-FASTCORR) and STandUP. D.T.~acknowledges support from the Swedish Research Council (VR) with grant No. 2019-03666. Y. L. acknowledges financial support from the KTH-CSC scholarship agreement (No.~201907090094). E.S.~acknowledges financial support from the Swedish Research Council (VR) through Grant No. 2017-03832.


\end{document}